\documentclass[camera,letterpaper,nomarginnotes,nonarrowgutter]{jpaper}
\usepackage{listings}
\usepackage{fancyhdr}

\usepackage{datetime}

\usepackage{cite}
\usepackage{amsmath,amssymb,amsfonts}
\usepackage{algorithmic}
\usepackage{graphicx}
\usepackage{textcomp}
\usepackage{xcolor}
\usepackage{tikz}

\usepackage[utf8]{inputenc}
\usepackage[T1]{fontenc}

\usepackage{booktabs} % For formal tables
\usepackage{setspace}
\usepackage[italic]{mathastext}
\usepackage{array}
\usepackage{titlesec}
\usepackage[normalem]{ulem}
\usepackage{multirow}
\usepackage{multicol}
\usepackage{color}
\usepackage[font={small,bf}]{caption}
\usepackage{float}
\usepackage[font={small,bf}]{subcaption}
\usepackage[linesnumbered,ruled]{algorithm2e}
\usepackage{makecell}

\usepackage{pifont}
\usepackage[]{microtype}
            
\usepackage{todonotes} % For margin notes
\usepackage{marginnote} 
\usepackage{pifont}
\let\marginpar\marginnote
\usepackage{titlesec}

\usepackage[bookmarks=true,breaklinks=true,colorlinks,linkcolor=black,citecolor=blue,urlcolor=black]{hyperref}
\definecolor{denim}{rgb}{0.08, 0.38, 0.74}
\definecolor{darkolivegreen}{rgb}{0.33, 0.42, 0.18}
\definecolor{dgreen}{rgb}{0.00, 0.75, 0.00}
\definecolor{darkpink}{rgb}{0.88, 0.28, 0.54}
\definecolor{forestgreen}{rgb}{0.0, 0.27, 0.13}
\definecolor{amber}{rgb}{1.0, 0.49, 0.0}
\definecolor{lightyellow}{rgb}{0.980, 0.956, 0.623}
\definecolor{lightblue}{rgb}{0.980, 0.956, 0.623}
\definecolor{darkamber}{rgb}{0.5, 0.19, 0.0}
\definecolor{dkgreen}{rgb}{0,0.6,0}
\definecolor{gray}{rgb}{0.5,0.5,0.5}
\definecolor{mauve}{rgb}{0.58,0,0.82}
\definecolor{lightmauve}{rgb}{0.68,0.4,0.92}
\definecolor{chocolate}{rgb}{0.48, 0.25, 0.0}
\definecolor{dollarbill}{rgb}{0.52,0.73,0.4}
\definecolor{dkdkgreen}{rgb}{0,0.45,0}
\definecolor{gfored}{rgb}{0.580, 0.050, 0.211}
\definecolor{darkwarmgray}{rgb}{0.15, 0.050, 0.05}
\definecolor{ups-truck}{rgb}{0.53, 0.28, 0.21}

\makeatletter
\g@addto@macro{\normalsize}{%
  \setlength{\abovedisplayskip}{2pt plus 1pt minus 1pt}
  \setlength{\belowdisplayskip}{2pt plus 1pt minus 1pt}
  \setlength{\intextsep}{2pt plus 1pt minus 1pt}
  \setlength{\textfloatsep}{3pt plus 1pt minus 1pt}
  \setlength{\dbltextfloatsep}{3pt plus 1pt minus 1pt}
  \setlength{\skip\footins}{4pt plus 1pt minus 1pt}
}
\def\BibTeX{{\rm B\kern-.05em{\sc i\kern-.025em b}\kern-.08em
    T\kern-.1667em\lower.7ex\hbox{E}\kern-.125emX}}

\def\UrlBreaks{\do\/\do-\/\do.\/\do:}

\expandafter\def\expandafter\UrlBreaks\expandafter{\UrlBreaks
  \do\a\do\b\do\c\do\d\do\e\do\f\do\g\do\h\do\i\do\j
  \do\k\do\l\do\m\do\n\do\o\do\p\do\q\do\r\do\s\do\t
  \do\u\do\v\do\w\do\x\do\y\do\z\do\A\do\B\do\C\do\D
  \do\E\do\F\do\G\do\H\do\I\do\J\do\K\do\L\do\M\do\N
  \do\O\do\P\do\Q\do\R\do\S\do\T\do\U\do\V\do\W\do\X
  \do\Y\do\Z}
  
\newcommand{\squishlist}{
 \begin{list}{$\circ$}
  { \setlength{\itemsep}{0pt}
     \setlength{\parsep}{0pt}
     \setlength{\topsep}{0pt}
     \setlength{\partopsep}{0pt}
     \setlength{\leftmargin}{1em}
     \setlength{\labelwidth}{1em}
     \setlength{\labelsep}{0.5em} } }

\newcommand{\squishsublist}{
\begin{list}{$\rightarrow$}
 { \setlength{\itemsep}{0pt}
    \setlength{\parsep}{0pt}
    \setlength{\topsep}{-10em}
    \setlength{\partopsep}{-3pt}
    \setlength{\leftmargin}{1em}
    \setlength{\labelwidth}{1em}
    \setlength{\labelsep}{0.5em} } }

\newcommand{\squishend}{
  \end{list}  }

\def\BibTeX{{\rm B\kern-.05em{\sc i\kern-.025em b}\kern-.08em
    T\kern-.1667em\lower.7ex\hbox{E}\kern-.125emX}}
\begin{document}

\title{Memory-Centric Computing}

\newcommand{\affilETH}[0]{\small {$$}}
\author{\vspace{-18pt}\\%
% \fontsize{11}{12}\selectfont%
{Onur Mutlu}%
% {Can Firtina}\quad%
\vspace{-3pt}\\%
% {\fontsize{10}{11}\selectfont
% \qquad\qquad\qquad\qquad\qquad\qquad
\affilETH\emph{ETH Z{\"u}rich}%
% \qquad\qquad
% \affilCMU\emph{Carnegie Mellon University}%
% }
\vspace{-12pt}}

\maketitle
\thispagestyle{plain}

\setstretch{0.79}

\begin{abstract}

Modern computing systems are processor-centric. Data processing (i.e., computation) happens only in the processor (e.g., a CPU, GPU, FPGA, ASIC). As such, data needs to be moved from where it is generated/captured (e.g., sensors) and stored (e.g., storage and memory devices) to the processor before it can be processed. The processor-centric design paradigm greatly limits the performance \& energy-efficiency, as well as scalability \& sustainability, of modern computing systems. Many studies show that even the most powerful processors and accelerators waste a large fraction (e.g., >60\%) of their time simply waiting for data and energy on moving data between storage/memory units to the processor. This is so even though most of the hardware real estate of such systems is dedicated to data storage and communication (e.g., many levels of caches, DRAM chips, storage systems, and interconnects).

Memory-centric computing aims to enable computation capability in and near all places where data is generated and stored. As such, it can greatly reduce the large negative performance and energy impact of data access and data movement, by fundamentally avoiding data movement and reducing data access latency \& energy. Many recent studies show that memory-centric computing can greatly improve system performance and energy efficiency. Major industrial vendors and startup companies have also recently introduced memory chips that have sophisticated computation capabilities.

This talk describes promising ongoing research and development efforts in memory-centric computing. We classify such efforts into two major fundamental categories: 1) processing using memory, which exploits analog operational properties of memory structures to perform massively-parallel operations in memory, and 2) processing near memory, which integrates processing capability in memory controllers, the logic layer of 3D-stacked memory technologies, or memory chips to enable high-bandwidth and low-latency memory access to near-memory logic. We show both types of architectures (and their combination) can enable orders of magnitude improvements in performance and energy consumption of many important workloads, such as graph analytics, databases, machine learning, video processing, climate modeling, genome analysis. We discuss adoption challenges for the memory-centric computing paradigm and conclude with some research \& development opportunities.

\end{abstract}

\vspace{-8pt}
\section{Memory-Centric Computing} \label{sec:introduction}
\vspace{-5pt}

Memory-centric computing (also called {\em processing in memory, PIM}) is a processing paradigm where data processing is performed near and in devices where data is generated (e.g., sensors) or stored (e.g., memory and storage devices)~\cite{mutlu2020modern}. This paradigm enables computing to be more efficient by offering an alternative to modern systems, which overwhelmingly use the processor-centric paradigm where data processing is performed {\em only} in the processor (which can be a CPU, GPU, FPGA, ASIC in modern systems). Memory-centric computing has several advantages over processor-centric computing. First, it fundamentally reduces the data movement bottleneck~\cite{boroumand_google_2018}, which plagues processor-centric systems that have to move data to the processor before processing it. Second, it enables low-latency and low-energy access to data by reducing the distance between processing units and data storage \& sensing units. Third, it can exploit large amounts of parallelism present in modern memory, storage, and sensor arrays to perform massively parallel (bit-level) computation~\cite{seshadri_dram_2020}. As such, memory-centric computing promises to improve both performance and energy-efficiency at the same time. 

Memory-centric computing systems can be categorized into two types~\cite{mutlu_processing_2019, mutlu2020modern}, based on the fundamental way in which computation is performed: 1) processing using memory (PuM), and 2) processing near memory (PnM). We briefly describe these next and give examples from recent works. These two approaches can be combined to obtain the best of both approaches. 

\vspace{-6pt}
\subsection{Processing using Memory (PuM)}
\vspace{-3pt}

A memory device has analog operational properties that enable it to perform (varying types and amounts of) computation. PuM exploits these properties to perform computation {\em using} the memory device (including memory cells, bitlines, wordlines, sensing structures, and peripheral circuitry). As such, the PuM approach can enable computation {\em without} adding logic to perform computation into a memory device, which makes it fundamentally different from modern processor-centric systems as well as PnM systems that add such logic near or in memory devices. PuM approach can be made more powerful by designing the memory device to increase its capability to perform analog computation.

% and data movement.  

PuM approaches have been demonstrated in DRAM (e.g.,~\cite{seshadri_rowclone_2013, seshadri_fast_2015, seshadri_buddy-ram_2016, seshadri_ambit_2017, seshadri_dram_2020, hajinazar_simdram_2021, pluto-micro22}), NVM (e.g.,~\cite{li2016, shafiee2016isaac, chi2016prime}), NAND flash (e.g.,~\cite{flashcosmos, parabit}) and SRAM (e.g.,~\cite{kang.icassp14, aga.hpca17}) devices. For example, recent works~\cite{seshadri_rowclone_2013, gao2020computedram, pidram} show that data copy and initialization can be performed inside a DRAM chip by exploiting internal connectivity in the DRAM chip, even in existing real DRAM chips that do {\em not} explicitly support these operations. Latency of a 4KB data copy can be improved by more than 11X and energy by 77X compared to a state-of-the-art processor-centric solution. Recent works~\cite{seshadri_fast_2015, seshadri_ambit_2017, gao2020computedram, hajinazar_simdram_2021} also show that bulk bitwise operations (Majority, AND, OR, NOT) and true random number generation~\cite{kim.hpca19, olgun2021quactrng} can be performed in commodity DRAM chips with small modifications or by violating timing parameters. Frameworks and compilers have been introduced to implement any type of operation using such bulk bitwise computation capability, with little effort required from the programmer~\cite{hajinazar_simdram_2021}. Real NAND flash memory chips can also perform bulk bitwise operations (AND, OR NOT, XOR) using inherent operational properties of NAND flash cells and strings as well as peripheral circuitry~\cite{flashcosmos, parabit}. Some emerging memory technologies are capable of performing matrix-vector multiplication operations in the analog domain due to their crossbar array structure~\cite{shafiee2016isaac, chi2016prime}, and various test chips have been designed to demonstrate this as proof-of-concept prototypes. 

\vspace{-6pt}
\subsection{Processing near Memory (PnM)}
\vspace{-3pt}

PnM adds processing logic (similar to modern processors and accelerators) close to or inside a memory device such that the distance between the processing logic and memory device is much smaller than in processor-centric systems. Such logic can be added to memory controllers, the logic layer of 3D-stacked memories, around peripheral circuitry in a memory chip, near memory subarrays in a memory chip, etc. The closer the logic is to the data storage parts of memory, the lower the amount of data movement. As such, PnM is not fundamentally different from modern systems where processing logic and memory structures are distinct, yet PnM greatly reduces the distance between them and in more aggressive implementations places logic and memory together in a tightly-integrated manner. 

Many recent works (e.g.,~\cite{ahn_scalable_2015, besta2021sisa, fernandez_natsa_2020, ghiasi2022genstore}) have shown the benefits of the PnM approach, by especially focusing on how various different types of applications can be accelerated using such an approach with varying levels of modifications to applications. For example, rewriting the entire application and changing the programming model to execute graph analytics near memory can greatly improve both performance and efficiency, by more than an order of magnitude~\cite{ahn_scalable_2015, besta2021sisa}. Less intrusive PnM approaches offload specific functions or instructions to near-memory logic~\cite{boroumand_google_2018, hsieh_transparent_2016, boroumand2021google, ahn.pei.isca15, hsieh_accelerating_2016, boroumand2021icde}, with lower but still large performance and energy benefits. 

\vspace{-6pt}
\subsection{Real PIM Systems}
\vspace{-3pt}

Recently, several real DRAM-based PnM systems were introduced as commercial systems or promising prototypes. The UPMEM company, for example, introduced a system where DRAM chips contain a general-purpose multithreaded processor next to each DRAM bank~\cite{devaux2019}. Several studies of the UPMEM system (e.g.,~\cite{gomez-luna_benchmarking_2021, gomez-luna_benchmarking_2022, giannoula2022sigmetrics, upmem-ml, diab2023bioinf-aim}) demonstrate the benefits and tradeoffs of this first commercial memory-centric system on various workloads and present benchmark suites and libraries for it. These studies show large performance and energy benefits when the workload is carefully designed to fit the constraints present in the PnM system, which is limited in terms of the computation power within the near-memory processors and the communication capability present between such processors and the host CPU. These studies also indicate how future general-purpose PnM systems can be improved to be much more powerful and effective. 

Several major vendors developed specialized PnM systems targeted toward machine learning applications and recommendation systems. For example, Samsung introduced FIMDRAM~\cite{kwon202125}, which is intended to accelerate floating-point based matrix operations (with native support for FP multiply and accumulates) in a DRAM chip. FIMDRAM incorporates processing units next to DRAM banks. To accelerate similar applications, SK-Hynix introduced the AiM-DRAM system~\cite{lee2022isscc}, which also incorporates near-bank computation units. Two other PnM systems were introduced by Alibaba~\cite{niu2022isscc} and Samsung/Meta~\cite{ke2021near} to accelerate recommendation systems. The former modifies a DRAM chip to perform specialized computation tailored towards recommendation inference. The latter includes a processing buffer chip in a DRAM module that performs similarly specialized computation on data coming from many DRAM chips surrounding it. 

Systems where computation can be offloaded to FPGAs that are equipped with high-bandwidth memory (HBM) also exist~\cite{singh_fpga-based_2021}. These systems can provide significant performance and energy benefits on various applications (e.g.,~\cite{singh_fpga-based_2021, singh_nero_2020}), including weather modeling and genome analysis. 

\vspace{-6pt}
\subsection{Adoption Challenges}
\vspace{-3pt}

Even though real PIM systems exist, memory-centric computing is far from being adopted in a widespread manner. To reach that level and thus realize the full potential and benefits of memory-centric computing, a number of challenges likely need to be solved. Many of these adoption challenges are common to PnM and PuM systems. We briefly cover some of these challenges, as they constitute important areas to investigate both in research and development of memory-centric computing systems.  

%, while PuM systems have some specific other challenges due to the fundamentally analog nature of computation they enable and the resulting bitwise-parallel compuation model they expose

First, it is important to accurately and comprehensively demonstrate which workloads and algorithms can benefit from memory-centric computing and by how much. This can enable a larger momentum for adopting PIM systems. It is especially critical to maximize benefits on important workloads. Second, widespread adoption of PIM requires such systems to be easy to program~\cite{ghoseibm2019}, which in turn requires support for seamless programming and compilation. Third, system and security support is needed to enable high efficiency and ease of use/programming. This wide topic includes support for data coherence between PIM and other computation units (e.g., CPUs)~\cite{boroumand_conda_2019, boroumand_lazypim_2016}, synchronization~\cite{giannoula_syncron_2021}, virtual memory~\cite{hsieh_accelerating_2016}, multiprogramming and sharing of PIM computation units, isolation between processes executing on PIM units, and communication interfaces to access PIM units. Fourth, it is important to design runtimes and compilation systems to decide what code should be executed in PIM units~\cite{oliveira_damov_2021}, how data should be mapped to facilitate PIM execution, and how access control and data sharing should be managed. Fifth, there is continual need for infrastructures and benchmarks (e.g.,~\cite{oliveira_damov_2021, pidram, gomez-luna_benchmarking_2022, giannoula2022sigmetrics, upmem-ml, ramulator, singh2019napel}) that help both hardware designers and software designers to accurately assess benefits, tradeoffs, and feasibility of different types of memory-centric computing systems. Finally, it is important to lower cost and demonstrate TCO benefits.

PuM systems have specific additional challenges due to their analog nature of computation. These include how to tolerate circuit variation and noise, how to ensure reliable operation, and how to enable computation on large memory arrays for scalable performance. In addition, some PUM systems implemented using memories that have endurance problems exacerbate lifetime and endurance problems. Due to such challenges, we believe PuM systems are harder to adopt in the short term even though their benefits can be fundamentally higher than PnM systems. 

\vspace{-6pt}
\subsection{Future Opportunities and Outlook}
\vspace{-3pt}

Memory-centric computing can enable balanced and efficient system designs where computation and memory access are fundamentally balanced and the processor-memory dichotomy is eliminated. These systems can provide greatly higher performance and efficiency than existing processor-centric systems. They can also enable potentially new applications and computing platforms. However, as with any new paradigm, memory-centric computing systems pose significant adoption challenges. We believe the processor-centric mindset that is ingrained in essentially every decision made in modern computing systems is likely the largest adoption challenge memory-centric systems face. We conclude that the future of memory-centric computing is very bright, but there is a lot more exciting research and development to do.

%\section*{Acknowledgments}

%We thank all members of the SAFARI Research Group for the stimulating and scholarly intellectual environment they provide. We acknowledge the generous gift funding provided by our industrial partners (especially by Google, Huawei, Intel, Microsoft, VMware). This work was in part supported by the Semiconductor Research Corporation (SRC) and BioPIM.

% \bibliographystyle{unsrt}
\setstretch{0.85}
% \setstretch{0.7}

\vspace{-8pt}
\bibliographystyle{IEEEtran}
\bibliography{IEEEabrv, main, safari_bio, safari_pim, references}
\end{document}